\documentclass[prd,nofootinbib]{revtex4}
\usepackage{epsfig,amsmath}

\begin{document}

\title{Remark on statistical model fits to particle ratios in relativistic
heavy ion collisions}

\author{F.~Becattini}\affiliation{Universit\`a di 
 Firenze and INFN Sezione di Firenze, Florence, Italy}

\begin{abstract}
In order to determine the chemical freeze-out parameters of the hadron-emitting 
source in relativistic heavy ion collisions some studies in literature perform
fits by using as data input a subsample of ratios calculated out of experimentally 
measured hadron yields instead of yields themselves. We show that this is a 
statistically incorrect method fit, implying a bias in the extracted parameters.
\end{abstract}

\maketitle

\section{Introduction}

The statistical model is succesfull in describing mean hadron multiplicities 
in both elementary and heavy ion collisions at high energy. In heavy ion 
collisions, many groups have carried out analysis of the available measurements 
for almost 10 years by now, reporting success in reproducing the data with few 
parameters. The usual employed technique is a $\chi^2$ fit, where the data is 
compared to the model prediction to determine the best values of those parameters 
(temperature, baryon-chemical potentials, volume and, possibly, non-equilibrium 
parameters $\gamma_S$ and $\gamma_q$). This analysis technique was first applied 
to heavy ion collision data in ref.~\cite{bgs}. 

The calculations performed by different groups agree to a very good degree of 
accuracy with the {\em same} data input, showing that the implementations of the 
statistical model are essentially consistent, modulo the small uncertainty
due to the poorly known branching ratios of high mass resonances. However, the 
parameter values quoted by different groups show significant differences, 
especially $\gamma_S$, which are ultimately owing to {\em different} data 
input. The main difference is related to the fact that some authors used hadron 
yields at midrapidity and others used full phase space yields, in the NN 
centre-of-mass energy range from few GeV to 17.2 GeV. This is due to different 
assumptions regarding the physical scheme, which can be settled after a careful 
study of the data. The point of view of the author of this note is expounded in 
ref.~\cite{beca1}, but we will not deal with this issue here. In fact, the main 
point of this note is to show that there is another source of discrepancy:
an incorrect fitting method which has led to biased results. The error resides
in the use, in the fits, of ratios of midrapidity or full phase 
space yields formed {\em a posteriori} out of quoted measurements. 

It should be stressed, from the very beginning, that this in incorrect procedure
only if the yields quoted by the experiments are used to form ratios 
{\em a posteriori} in a subsequent fit analysis. Experiments (especially at RHIC) 
often quote ratios of yields (e.g. $K^+/K^-$ or $\bar p/p$) because they can thereby
achieve some nice cancellation of systematic errors. In this case, ratios are measured 
by experiments and using them in a fit is perfectly legitimate and sound; what is 
wrong is getting measured yields and replace them with ratios made ``at home". 
 
\section{Why not to form particle ratios}

When trying to determine the best fit parameters of the statistical model in relativistic
heavy ion collisions, one 
is given particle multiplicities or ratios of multiplicities, either at midrapity 
(i.e. $dN/dy$ for $y = 0$ in the centre-of-mass frame) or in full phase space. As
has been mentioned, the 
experiments sometimes quote ratios because of some cancellation of systematic errors, 
so that the effective error on the ratio is consistently smaller than 
that one would get if the yields had uncorrelated errors. In most cases, though, 
instead of ratios, experiments quote yields, which were obtained by 
means of extrapolating fits to both $p_T$ and rapidity spectra. Unless some definite 
information is provided by the experimentalists, one is supposed to assume the errors 
on the yields to be fully uncorrelated, so that the no further information can be 
obtained by just manipulating the yields {\em a posteriori}. However, some authors 
have formed ratios out of quoted experimental multiplicities and replaced yields 
with an {\em equal amount} (or diminished by one) of ratios in the statistical model
fits to determine temperature and baryon-chemical potential, even without taking 
into account relevant correlations between different ratios. This procedure is meant 
to get rid of the volume parameter in the fit, which is an overall normalization factor 
cancelling out in ratios, leaving only the intensive parameters $T$, $\mu_B$ and 
$\gamma_S$. However, this method would be incorrect even if correlations were 
taken into account and, most likely, involves a bias in the determination of the parameters 
themselves. We will illustrate how this problem comes about by first giving three 
simple examples and presenting a realistic one in next section 2.

\subsection{Example 1}

The simplest example is the weighted average. Consider four independent measurements
of the same quantity $x$ with different normal errors, say $x_1 = 1.2 \pm 0.2$, 
$x_2 = 0.8 \pm 0.2$, $x_3 = 0.8 \pm 0.2$, $x_4 = 0.8 \pm 0.2$ \footnote{In statistics,
strictly speaking, the validity of the presently described $\chi^2$ minimizations and 
tests involving normally distributed random variables requires the errors to be the 
{\em true} variances \cite{roe}. We will assume throughout the paper that this is the 
case, i.e. we assume that our errors are the actual value of the gaussian parameter 
$\sigma$}. It is well known that the problem of determining the best 
estimate of $x$ through maximum likelihood method leads to the minimization of the $\chi^2$:
\begin{equation}
 \chi^2 = \sum_{i=1}^4 \frac{(x_i - x)^2}{\sigma_i^2} 
\end{equation}
and has the weighted average as solution, which is in this case $\bar x = 0.9 \pm 
0.1$ with a $\chi^2/dof = 1$, that is a very good fit. If, on the other hand, we 
want to assess the consistency of the four measurements by taking ratios of pairs, 
we soon face an ambiguity: how many ratios should one take? The naive (wrong) answer is 
taking as many as degrees of freedom in the $\chi^2$ minimization, that is 3 in our 
example. Yet, there are 6 different triplets ($N(N-1)/2$ in general) which can be 
formed out of 4 objects, considering as equivalent a ratio $x_i/x_j$ and its inverse
$x_j/x_i$. Therefore, a choice has to be made; for instance, if we took $x_1/x_2$, 
$x_1/x_3$ and $x_1/x_4$, we would get three times 1.5, whereas if we took $x_3/x_2$,
$x_3/x_4$ and $x_2/x_1$, we would get 1.0 twice and 0.66. The two triplets of ratios 
(1.5,1.5,1.5) and (1.0,1.0,0.66) submitted to a consistency test yield different 
answers in terms of statistical significance, even taking into account the correlations 
between them. The deep reason of this is an information loss in using a subsample
of ratios of measurements instead of measurements themselves; by retaining only three 
ratios out of six to naively avoid redundancy, one is forced to give up some information 
and the statistical significance does depend on the particular chosen subset 
of ratios. 

\subsection{Example 2}

In this example we provide a concrete numerical example demonstrating the involved error 
in fitting a subsample of ratios instead of measurement. Consider a simple 
linear model $y = x + c$ to be fitted to the measurements $(x, y) = (1, 2.3 \pm 0.1), 
(2, 2.8 \pm 0.13), (3, 4.3 \pm 0.08)$. The correct fit procedure to determine the best 
value of $c$ is the minization of:
\begin{equation}\label{correct}
 \chi^2 = \sum_{i=1}^3 \frac{(y_i - x_i - c)^2}{\sigma_i^2} 
\end{equation}
If, on the other hand, we adopted the ratio method, we should have chosen two ratios formed 
out of the three measurements, e.g. $y_1/y_3$ and $y_2/y_3$ and minimize the $\chi^2$:
\begin{equation}\label{wrong}
 \chi^2 = \sum_{i,j=1}^2 \left(R_i - \frac{x_{k(i)} + c}{x_{h(i)} + c} \right) 
 C^{-1}_{ij}\left(R_j - \frac{x_{k(j)} + c}{x_{h(j)} + c} \right)
\end{equation}
where $k(i)$, $h(i)$ are the indices of the measurements used to form the $i$th 
ratio and $C$ is the covariance matrix with non-vanishing off-diagonal elements 
estimated by means of the error propagation rules. For this example, the 
$\chi^2$ profiles as a function of $c$ are compared in fig. 1. One can clearly 
see that both the minimum and the curvature of the $\chi^2$ around the minimum are 
different for the two functions (\ref{correct}) and (\ref{wrong}). This is reflected 
in different estimates of the best fit value and its error. It is especially worth 
remarking  that the error estimate related to (\ref{wrong}) is much larger than the 
correct error.

\begin{center}
\begin{figure}[htb]
\epsfxsize=4.0in
\epsffile{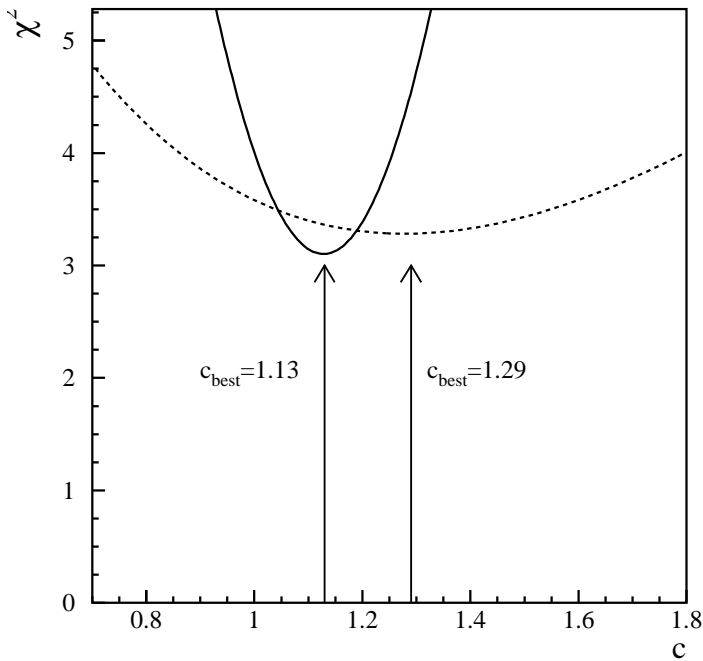} 
\caption{$\chi^2$ profile for the example 2}
\end{figure}
\end{center} 

\subsection{Example 3}

We are now going to consider an example which is quite close to the actual problem of 
fitting multiplicities to the statistical model, which will make it clear that 
replacing $N$ multiplicities with $N-1$ ratios is incorrect also for an 
exponential fit like the statistical-thermal model one. Consider a model $y = a \exp(bx)$ 
and three measurements $(x, y) = (0, 1.8\pm0.1), (1, 2.71\pm 0.13), (2, 6.5 \pm 0.08)$. 
Here, the parameter $a$ corresponds to a volume and $b$ to an inverse temperature. 
The idea is to get rid of the parameter $a$ and fitting just $b$ by taking ratios of 
measurements. The correct $\chi^2$ now reads: 
\begin{equation}\label{correct3}
 \chi^2 = \sum_{i=1}^3 \frac{(y_i - a \exp(bx_i))^2}{\sigma_i^2} 
\end{equation}
whereas the $\chi^2$ for the ratios $R_1=y_1/y_3$, $R_2=y_2/y_3$ reads:
\begin{equation}\label{wrong3}
 \chi^2 = \sum_{i,j} \left(R_i - \exp[b(x_{k(i)}-x_{h(i)})] \right) 
 C^{-1}_{ij} \left(R_j - \exp[b(x_{k(j)}-x_{h(j)})]\right)
\end{equation}
The parameter $a$ has disappeared in eq.~(\ref{wrong3}). If we have a look at $\chi^2$ 
profiles of (\ref{correct3}) and (\ref{wrong3}) as a function of $b$, we can see that 
also in this case both minima and curvature around the minima differ. The difference 
between the correct estimate of $b$ and the one obtained by minimizing (2.5) is 2\% 
which is little but cannot be neglected aiming at reaching the best accuracy.
\\
\begin{center}
\begin{figure}[htb]
\epsfxsize=4.0in
\epsffile{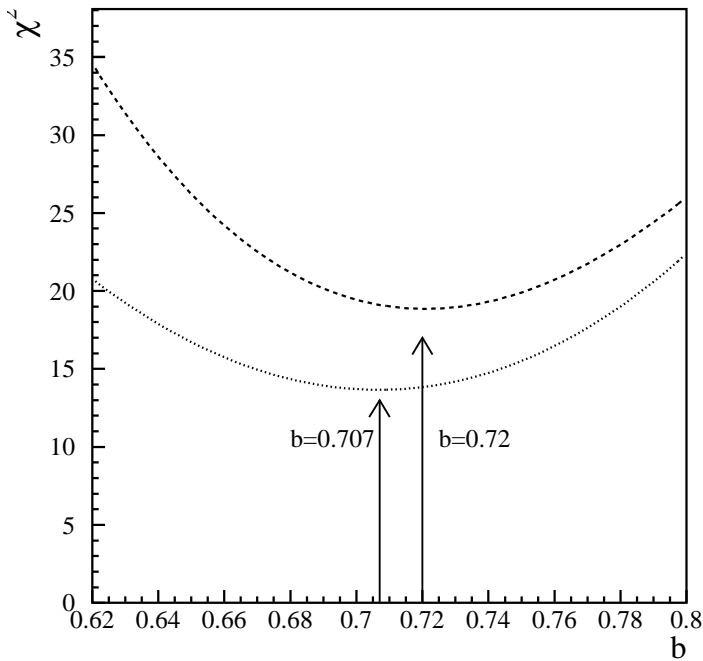} 
\caption{$\chi^2$ profile for the example 3}
\end{figure}
\end{center} 

From the previous three examples we have learned how the replacement of measurements 
with ratios of them, always involves a discrepancy with the results of the maximum 
likelihood method. Since a fundamental theorem of statistical inference states that 
if an efficient unbiased estimator exists, the maximum likelihood method will find it, 
we can fairly conclude that the estimate obtained with the minimization of the 
pseudo-$\chi^2$'s with ratios of normally distributed random variables is not efficient
and biased. Although we do not provide a general analytical proof of this statement, 
example 2 clearly shows that the estimator is not efficient as the curvature of the
$\chi^2$ in the minimum (which, for a linear fit, is related to the inverse variance 
of the estimator) is larger for the correct method than for the ``ratios" method.
Thus, most likely, bias and inefficiency occur on a more general ground.

\section{An example at RHIC}

We will now provide a realistic example of how the use of ratios instead of yields 
alters the estimation of thermodynamical parameters. We will use data collected by STAR 
experiment at RHIC, at $\sqrt s_{NN} = 130$ GeV, shown in table 1. We stress that 
experimental numbers here serve just as an instrument to compare different fit 
procedures; most probably, some ratio measured by STAR experiment has a smaller 
systematic error than yields and should be better used (for a more exhaustive 
discussion on how they have been collected and other comments, see ref.~\cite{jaakko}). 
Nevertheless, this is not the point here: the aim of this exercise is to show how 
the choice of a particular set of ratios can affect the fit outcome. In this 
particular case, we have to pick 11 ratios out of 66 and we then have much freedom. 
Let us first perform a correct fit to the midrapidity yields, 
as quoted in table 1, and construct an array of residuals, i.e. a set of differences 
between actual measurements and fitted values, divided by the experimental error. 
Indeed, it is fairly easy to realize that if we systematically choose ratios of 
light particles with negative residual to heavy particles with positive residual, 
the ensuing value of $T$ from a new fit to the ratios will tend to be larger, implying
that we have introduced a bias in the fit. To show this, we have first 
performed a fit to midrapidity densities in table 1 by fixing the strangeness 
suppression factor $\gamma_S$ to 1. We then had a look at residuals of all particles and 
took the heaviest particle with positive residual, i.e. $\bar\Xi^+$ . We then chose 
a set of 11 ratios $\langle X \rangle / \langle \bar\Xi^+\rangle$ , $X$ being any
particle lighter than $\Xi$ in table 1, and added as last ratio in the data 
sample $\langle p \rangle/ \langle \Omega + \bar\Omega \rangle$, also showing a 
fluctuation in the same direction. As has been mentioned, the expectation for this 
kind of analysis is to artificially enhance the temperature in this fit, because 
all lighter particles have a residual larger than $\bar\Xi^+$'s. This is indeed what is 
found, as shown in table 2. If we do not include correlations, the difference 
between fit to ratios and correct fit to the yields is 8.2 MeV, i.e. about 2.61$\sigma$.
The situation is much worse for $\mu_B$, with a difference of 13 MeV and $21\sigma$, 
which is owing to the same problem: choosing an upward fluctuating antibaryon as reference
implies the decrease of fitted $\mu_B$. On the other hand the situation
improves a lot including correlations (what is not anyway done in many analyses). 
Anyhow, the method is still conceptually wrong.

\begin{table}
\caption{List of midrapidity yields of different hadrons measured by STAR in Au-Au
collisins at $\sqrt s_{NN} = 130$ GeV compared with their fitted values.}
\vspace{0.5cm}
\begin{tabular}{|c|c|c|}
\hline
 Particle      & $dN/dy|_{y=0}$ measured & $dN/dy|_{y=0}$ fitted \\
\hline				    	  
 $\pi^+$       & $239\pm10.6$    &         237.1		  \\
 $\pi^-$       & $239\pm10.6$    &         240.0		  \\
 $K^+$         & $45.8\pm6.7$    &   	    45.8		  \\
 $K^-$         & $43.2\pm6.0$    &   	    42.5		  \\
 $p$           & $26.2\pm6.0$    &   	    35.6		  \\
 $\bar p$      & $18.9\pm4.3$    &   	    24.6		  \\
 $\Lambda$     & $17.2\pm1.8$    &   	    16.4		  \\
 $\bar\Lambda$ & $12.3\pm1.3$    &   	    12.2		  \\
 $\phi$        & $6.09\pm0.85$   &   	     6.14		  \\
 $\Xi^-$       & $2.13\pm0.27$   &   	     1.92		  \\
 $\bar\Xi^+$   & $1.78\pm0.24$   &   	     1.53		  \\
 $\Omega+\bar\Omega$ & $0.586\pm0.128$ &     0.642		  \\ 
\hline
\end{tabular}
\end{table}

\begin{table}
\caption{Results of statistical model fits to hadron midrapidity densities in
table 1. The first column shows the results of a fit to the set of ratios described
in the text. The third column shows the results of a fit to the same set of ratios
taking into account correlations in the $\chi^2$.}
\vspace{0.5cm}
\begin{tabular}{|c|c|c|c|}
\hline
 Parameters     & Fit to yields   & Fit to ratios w/o correlations & Fit to ratios with correlations\\
\hline
 T (MeV)        &$168.4\pm 3.2$   &$175.6 \pm 3.0$                 &$170.3 \pm 3.0$  \\
$\mu_B$(MeV)    &$36.2 \pm 0.6$   &$23.4  \pm 0.4$                 &$33.1 \pm 0.6$  \\
\hline
 $\chi^2/dof$   & 6.3/10          &  6.6/10                        &  7.6/10 \\
\hline
\end{tabular}
\end{table}

\section{Conclusions}

We have shown that replacing, {\em a posteriori}, measured yields with an equally
sized sample of particle ratios in statistical model (as well as in any other model)
fits to determine the chemical freeze-out parameters in relativistic heavy ion collisions 
is a statistically incorrect method. The error resides in the fact that selecting a 
subsample of all possible ratios entails a loss of information and produces biased and 
inefficient estimators. Only for vanishing measurement errors the two methods would 
be equivalent. 

We have shown, in a realistic example, that the difference between correct and wrong fit
may be as large as ${\cal O}(10)$ MeV for temperature and baryon-chemical potential if, 
as it is usually done, correlations are not taken into account. This difference has been
found with a purposely chosen set of ratios, designed to enhance the effect and it is
then unlikely that all analyses performed so far suffer from such a large bias. It
would be anyway appropriate to avoid introducing this bias in analyses aimed at 
getting the best accuracy.

\section*{Acknowledgments}

I would like to express my gratitude for clarifying discussions about 
statistical methods of data analysis to my colleagues in the Department 
of Physics of the University of Florence.



\begin{thebibliography}{99}
\section*{References}

\bibitem{bgs} 
 F. Becattini, M. Gazdzicki and J. Sollfrank, Eur. Phys. J. C 5 (1998) 143. 
\bibitem{beca1}  
  F.~Becattini, M.~Gazdzicki, A.~Keranen, J.~Manninen and R.~Stock,
  Phys. Rev. C 69 (2004) 024905.
\bibitem{roe} see e.g. B. Roe, {\it Probability and Statistics in Experimental 
 Physics}, Springer 1992
\bibitem{jaakko}
 J. Manninen, talk given at {\it Critical point and onset of deconfinement},
 Florence, July 3-6 2006, proceeding published in http://pos.sissa.it.

\end{thebibliography}
\end{document}